# Resonance properties of forced oscillations of particles and gaseous bubbles in a viscous fluid at small Reynolds numbers

H. K. Hassan[a)] and Y. A. Stepanyants[b)]

*University of Southern Queensland, West St., Toowoomba, QLD, 4350, Australia*

We consider small oscillations of micro-particles and gaseous bubbles in viscous fluid around equilibrium states under the action of a sinusoidal external force. Exact solutions to the governing integro-differential equations containing both Stokes and memory-integral drag forces are obtained. The main aim of this study is to clarify the influence of the memory-integral drag force on the resonance characteristics of oscillating particles or gaseous bubbles in a viscous fluid at small Reynolds numbers. The resonant curves (amplitude versus frequency of external force), as well as phase-frequency dependences are obtained for both these objects and compared with the corresponding dependences of the traditional oscillator with the Stokes drag force only.

---

[a)] Electronic mail: Hassanahlf@yahoo.com

[b)] Electronic mail: Yury.Stepanyants@usq.edu.au

As is well known[1], a small fluid drop moving in a viscous fluid at a small Reynolds number experiences an influence of at least two drag forces: one of them is the traditional Stokes drag (SD) force and another is the memory-integral drag (MID) force (the latter is also known as the Boussinesq–Basset drag (BBD) force in the case when the drop reduces to a solid particle). In the past decade a vast number of papers devoted to the role of MID force in the dynamics of solid particles, gaseous bubbles and other liquid drops in viscous fluids were published (see, e.g., Refs. [2–9] and references therein). The growing interest in this problem in recent years is associated with the development of a new field of microfluids and the technology of using micro- and nano-particles. Such technology is already used in medicine (for diagnostics, drug delivery to specific organs), biology, food quality control, chemistry, etc. (see the review Ref. [10] and references therein).

In many cases nanoparticles can experience an oscillatory motion around quasi-stationary positions under the action of external forces, for example, acoustic radiative force[6,7,9,10]. In such cases it is important to know the resonance properties of nanoparticles, e.g., the shape of the resonance curve (the dependence of amplitude of oscillation on the frequency of external sinusoidal force), width and amplitude of a resonance curve, the quality of the effective oscillator. All these characteristics are well known for the usual linear oscillator (see, e.g., Refs. [11, 12]) which is equivalent to the nanoparticle oscillator under the influence of Stokes drag force only. However, to our best knowledge the influence of the MID force on the resonant property of an oscillator has not been studied yet. Here we fill this gap for two limiting cases, when the oscillating drop is (i) a solid particle and (ii) a gaseous bubble.

Consider first small oscillations of a solid spherical particle of density $\rho_p$ and radius $a$ in a viscous fluid. Particle motion is caused by the influence of external harmonic force having the amplitude $\tilde{A}$ and frequency $\tilde{\omega}$. A corresponding equation of motion in a one-dimensional case in the creeping flow regime is [1, 5, 13]:

$$\left(r+\frac{1}{2}\right)\frac{d^2z}{dt^2}+\frac{9}{2}\frac{\mu}{a^2\rho_f}\frac{dz}{dt}+\frac{9}{2a}\sqrt{\frac{\mu}{\pi\rho_f}}\int_{-\infty}^{t}\frac{d^2z(\tau)}{d\tau^2}\frac{d\tau}{\sqrt{t-\tau}}+\tilde{\omega}_0^2 z=\tilde{A}\sin\tilde{\omega}t, \qquad (1)$$

where $z$ is the particle coordinate; $r\equiv\rho_p/\rho_f$ is the ratio of particle to fluid density; $\mu$ is the dynamic viscosity of a fluid, and $\tilde{\omega}_0$ is the frequency of free oscillations of a particle in the absence of dissipation ($\nu = 0$). In this equation the second term in the left-hand side describes the quasi-stationary Stokes drag (SD) force, while the integral term describes the well-known BBD force [1, 14]. The added mass effect for a spherical particle is accounted for through the coefficient 1/2 within the bracket in the left-hand side of equation [14]. We assume that a particle being at rest commences with an instantaneous motion at $t = 0$ with the initial velocity $V_0$, i.e. its velocity experiences a sudden jump from zero to $V_0$ and then varies in accordance with the equation of motion (1). Thus, the particle velocity, including the initial jump, can be expressed through the unit Heaviside function H($t$): $(dz/dt)_{tot} = H(t)(dz/dt)_{pos}$, where $(dz/dt)_{tot}$ is the velocity at any instant of time, whereas $(dz/dt)_{pos}$ relates to the velocity at positive times only. Correspondingly, the acceleration is $(d^2z/dt^2)_{tot} = \delta(t)(dz/dt)_{pos} + H(t)(d^2z/dt^2)_{pos}$, where $\delta(t) \equiv d\,H(t)/dt$ is the Dirac delta-function. Taking into consideration the effect of the Dirac delta-function under the integral, Eq. (1) for $t > 0$ can be presented in the form (the index "pos" has been omitted):

$$\left(r+\frac{1}{2}\right)\frac{d^2z}{dt^2}+\frac{9\mu}{2a^2\rho_f}\left[\frac{dz}{dt}+\frac{aV_0}{\sqrt{\pi\nu t}}+a\sqrt{\frac{\rho_f}{\pi\mu}}\int_{0+}^{t}\frac{d^2z(\tau)}{d\tau^2}\frac{d\tau}{\sqrt{t-\tau}}\right]+\tilde{\omega}_0^2 z=\tilde{A}\sin\tilde{\omega}t. \qquad (2)$$

The equation of motion can be expressed in the dimensionless form by introducing the following normalised variables: $\zeta = z/a$, $\theta = 9\mu t/a^2\rho_f$, $\upsilon_0 = V_0 a\rho_f/9\mu$. Equation (2) after that reduces to [5, 13]:

$$\frac{d^2\zeta}{d\theta^2}+\alpha\frac{d\zeta}{d\theta}+\frac{\beta}{\sqrt{\pi}}\left(\frac{\upsilon_0}{\sqrt{\theta}}+\int_0^{\theta}\frac{d^2\zeta(\vartheta)}{d\vartheta^2}\frac{d\vartheta}{\sqrt{\theta-\vartheta}}\right)+\omega_0^2\zeta=A\sin\omega\theta, \qquad (3)$$

where $\alpha = 1/(1 + 2r)$, $\beta = 1/(1 + 2r)$, $\omega_0 = \tilde{\omega}_0 \sqrt{2a^3 \rho_f^2 / 81\mu^2 (2r+1)}$, $\omega = \tilde{\omega} a^2 \rho_f / 9\mu$, $A = 2\tilde{A} a^3 \rho_f^2 / 81\mu^2 (2r+1)$. Although the coefficient $\alpha$ and $\beta$ are equal, we denote them by different letters; this allows us to switch off either the SD force setting $\alpha = 0$ or BBD force setting $\beta = 0$. We assume further that the viscosity of an ambient fluid is relatively small, so that the frequency of free oscillations $\omega_0 \gg \alpha, \beta$ (however the Reynolds number $\text{Re} = aV\rho_f/\mu \ll 1$ is still very small; this can be achieved, for example, by the smallness of particle radius $a$).

As this is a linear equation with respect to $\zeta$, its general solution consists of two parts, the general solution of homogeneous equation and a particular solution of non-homogeneous equation. The general solution of a homogeneous equation with $A = 0$ can be readily obtained with the help of the Laplace transform subject to the initial conditions $\zeta(0) = \zeta_0$ and $\dot{\zeta}(0) = \upsilon_0$ (here the dot on the top of a letter stands for a derivative with respect to $\theta$); this has been done in Refs. [5, 13]. We do not reproduce here the lengthy solution from those papers; it has been analysed there in detail and represents slowly decaying oscillations (provided that the decay coefficients $\alpha$ and $\beta$ are small compared to $\omega_0$).

The most important issue to us is the solution of the forced equation (3) with $A \neq 0$. To construct such a solution we choose a trial solution in the form $\zeta(\theta) = B\cos\omega\theta + C\sin\omega\theta$ and assume for simplicity that the particle is at the rest at the initial instant of time, so that $\zeta_0 = \upsilon_0 = 0$. With the trial solution chosen above the integral term in Eq. (3) can be readily evaluated (see also Ref. [6])

$$\int_0^\theta \frac{d^2\zeta(\vartheta)}{d\vartheta^2} \frac{d\vartheta}{\sqrt{\theta - \vartheta}} = -\frac{\omega^{3/2}}{2}\sqrt{\frac{\pi}{2}}\left[B(\sin\omega\theta + \cos\omega\theta) + C(\sin\omega\theta - \cos\omega\theta)\right]. \tag{4}$$

Substituting then the trial solution into Eq. (3) and using Eq. (4), we obtain after some manipulations

$$B = -\frac{A}{6}\omega \frac{2\alpha + 3\beta\sqrt{2\omega}/2}{\left(\omega_0^2 - \omega^2 - \sqrt{2}\beta\omega^{3/2}/4\right)^2 + \left(\alpha + 3\beta\sqrt{2\omega}/4\right)^2 \omega^2/9}, \tag{5}$$

$$C = A \frac{\omega_0^2 - \omega^2 - \beta\sqrt{2}\omega^{3/2}/4}{\left(\omega_0^2 - \omega^2 - \sqrt{2}\beta\omega^{3/2}/4\right)^2 + \left(\alpha + 3\beta\sqrt{2\omega}/4\right)^2 \omega^2/9}. \tag{6}$$

Let us present now the forced solution in the equivalent form

$$\zeta(\theta) = B\cos\omega\theta + C\sin\omega\theta = Amp(\omega_0, \omega)\sin(\omega\theta + \varphi), \tag{7}$$

where the amplitude of forced oscillations is

$$Amp(\omega_0, \omega) = \sqrt{B^2 + C^2} = \frac{A}{\sqrt{\left(\omega_0^2 - \omega^2 - \sqrt{2}\beta\omega^{5/2}\right)^2 + \frac{1}{9}\omega^2\left(\alpha + 3\sqrt{2}\beta\omega^{3/2}\right)^2}}, \tag{8}$$

and the phase $\varphi$ is determined from the equation

$$\tan\varphi = \frac{B}{C} = \frac{\omega}{12} \frac{2\alpha + 3\beta\sqrt{2\omega}}{\omega_0^2 - \omega^2 - \left(\sqrt{2}/4\right)\beta\omega^{3/2}}. \tag{9}$$

These expressions can be presented in the traditional dimensionless forms if we normalise the frequency $\nu = \omega/\omega_0$ and introduce the quality factors, $Q_\alpha = \omega_0/\alpha$ and $Q_\beta = 8\omega_0/\beta^2$. The quality factor $Q_\alpha$ is well known in the theory of oscillations (see, e.g., Ref. [12]), whereas another quality factor $Q_\beta$ is new. Taking into account our definitions of coefficients $\alpha$ and $\beta$ (see after Eq. (3)), we can express the quality factors in terms of frequency $\omega_0$ and relative particle density $r$: $Q_\alpha = \omega_0(1 + 2r)$ and $Q_\beta = 8\omega_0(1 + 2r)^2$. Normalising then the amplitude of particle oscillation $A_n = \omega_0^2 Amp/A$, we finally present the amplitude-frequency dependence (8) in the form

$$A_n(\nu, Q_\alpha, Q_\beta) = \frac{1}{\sqrt{\left(1 - \nu^2 - \nu\sqrt{\nu/Q_\beta}\right)^2 + \nu^2\left(1/Q_\alpha + \sqrt{\nu/Q_\beta}\right)^2}}. \tag{10}$$

And the phase-frequency dependence (9) in the dimensionless form reads

$$\varphi(v, Q_\alpha, Q_\beta) = \tan^{-1}\left(\frac{v}{Q_\alpha} \frac{1 + Q_\alpha \sqrt{v/Q_\beta}}{v^2 - 1 + v\sqrt{v/Q_\beta}}\right). \tag{11}$$

If the BBD force is neglected ($Q_\beta \to \infty$), then the dependences (10) and (11) naturally reduce to the classic ones[11,12]

$$A_n(v, Q_\alpha) = \left[(1-v^2)^2 + v^2/Q_\alpha^2\right]^{-1}, \quad \varphi(v, Q_\alpha) = \tan^{-1}\left[v/Q_\alpha(v^2-1)\right]. \tag{12}$$

Graphics of dependences of $A_n(v, Q_\alpha, Q_\beta)$ and $\varphi(v, Q_\alpha, Q_\beta)$ as functions of normalised frequency $v$ are shown in Figs. 1a) and 1b) respectively. Lines 1 pertain to the case when only the Stokes drag force is taken into account and the BBD force is ignored. The parameter $Q_\alpha = 700$ was chosen which corresponds to $\omega_0 = 100$ and $r = 3$ in Eq. (3). Lines 2 pertain to the case when both Stokes and BBD forces are taken into account with the same parameters $\omega_0$ and $r$ (this gives $Q_\beta = 39{,}200$).

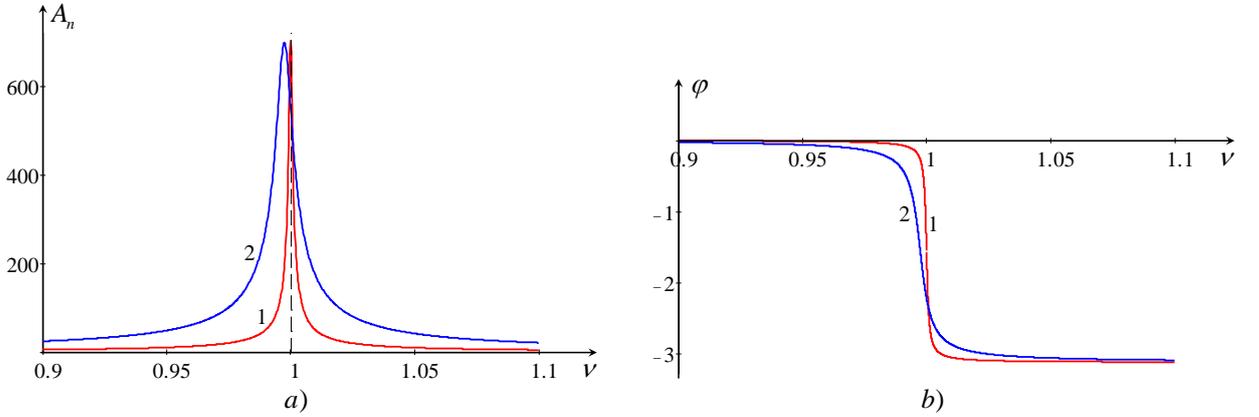

FIG. 1 (color online). The amplitude-frequency dependence (a) and phase-frequency dependence (b) for oscillating solid particle in a viscous fluid under the influence of only the Stokes drag force (lines 1) and both Stokes and BBD drag forces (lines 2). Function $A_n$ for line 2 in frame (a) was multiplied by the factor 4.5.

Thus, from this analysis one can see that under the influence of BBD force the maximum of the resonance curve becomes about 4.5 times smaller and correspondingly 4.5 times wider. It is also shifted to the left from $\nu = 1$ (see Fig. 1a).

Consider now small oscillations of a gaseous spherical bubble of a density $\rho_b$ and a radius $a$ in a viscous fluid. The bubble sphericity can be maintained by a surface tension which is very important for small-radius bubbles. The basic equation of motion in the dimensionless variables is similar to Eq. (3), but contains different coefficients of viscous terms and the different kernel of MID force (the same dimensionless variables can be used in this case too). Now instead of the Stokes drag force the corresponding term proportional to the first derivative $d\zeta/d\theta$ is known as the Hadamard–Rybczynski drag (HRD) force (see, e.g., Refs. [14, 5]), and the basic equation reads (its derivation is similar to that presented above for a solid particle):

$$\frac{d^2\zeta}{d\theta^2} + \frac{2}{3}\alpha \frac{d\zeta}{d\theta} + \frac{4}{3}\beta\left(\frac{\upsilon_0}{\sqrt{\theta}} + \int_0^\theta \frac{d^2\zeta(\vartheta)}{d\vartheta^2} e^{\theta-\vartheta}\mathrm{erfc}\left(\sqrt{\theta-\vartheta}\right)d\vartheta\right) + \omega_0^2 \zeta = A\sin\omega\theta. \quad (13)$$

Here erfc $(x) = 1 − $ erf $(x)$ is the complimentary error function[15], and $\alpha = \beta = 1$, because the ratio of bubble density to the fluid density is negligible, $r = \rho_b/\rho_f \approx 0$. However, we will keep again the parameters $\alpha$ and $\beta$ as arbitrary to follow up the influence of the HRD and MID forces separately.

Assuming again the zero initial conditions and focussing on the steady oscillations of a bubble after a transient process, we choose a trial solution in the form $\zeta(\theta) = B\cos\omega\theta + C\sin\omega\theta$. Evaluate first the integral term in Eq. (13) with the trial solution

$$I(\theta) = \int_0^\theta \frac{d^2\zeta(\vartheta)}{d\vartheta^2} e^{\theta-\vartheta}\mathrm{erfc}\left(\sqrt{\theta-\vartheta}\right)d\vartheta = -\omega^2\int_0^\theta \left(B\cos\omega\vartheta + C\sin\omega\vartheta\right) e^{\theta-\vartheta}\mathrm{erfc}\left(\sqrt{\theta-\vartheta}\right)d\vartheta. \quad (14)$$

To evaluate this integral, consider its Laplace transform

$$\hat{I}(s) = -\omega^2 \frac{Bs + C\omega}{s^2 + \omega^2} \frac{1}{\sqrt{s}\left(\sqrt{s}+1\right)} = -\omega^2 \frac{Bs + C\omega}{s^2 + \omega^2}\left[\frac{1}{s-1} - \frac{\sqrt{s}}{s(s-1)}\right]. \quad (15)$$

This expression can be further presented in terms of a sum of simplest ratios

$$\hat{I}(s) = -\frac{\omega^2}{\omega^2+1}\left[(B\omega-C)\frac{\omega}{s^2+\omega^2} - (B+C\omega)\frac{s}{s^2+\omega^2} + (B+C\omega)\frac{1}{s-1} + \frac{C}{\omega}\frac{\omega^2+1}{\sqrt{s}}\right.$$

$$\left. -(B+C\omega)\frac{\sqrt{s}}{s-1} + (B+C\omega)\frac{\sqrt{s}}{s^2+\omega^2} + \frac{B\omega-C}{\omega}\frac{s\sqrt{s}}{s^2+\omega^2}\right]. \tag{16}$$

The inverse Laplace transform then gives

$$I(\theta) = -\frac{\omega^2}{\omega^2+1}\left\{(B+C\omega)e^{\theta}\left[1-\text{erf}\left(\sqrt{\theta}\right)\right] + (B\omega-C)\sin\omega\theta - (B+C\omega)\cos\omega\theta\right.$$

$$\left. -\frac{\sqrt{i}}{2\omega}e^{i\omega t}\text{erf}\left(\sqrt{i\omega t}\right)\left[B\omega-C+i\sqrt{\omega}(B+C\omega)\right] - \frac{\sqrt{-i}}{2\omega}e^{-i\omega t}\text{erf}\left(\sqrt{-i\omega t}\right)\left[B\omega-C-i\sqrt{\omega}(B+C\omega)\right]\right\}. \tag{17}$$

The first term in the curly brackets $e^{\theta}[1 - \text{erf}(\sqrt{\theta})]$ quickly vanishes when $t \to \infty$. All other terms after simplification reduce to

$$I(\theta) = -\omega(B\sin\omega\theta - C\cos\omega\theta). \tag{18}$$

Substituting then the trial solution into Eq. (13) and using Eq. (18), we obtain after simple manipulations

$$B = -\frac{2A}{9}\omega\frac{\alpha+6\beta}{\left(\omega^2-\omega_0^2\right)^2 + 4(\alpha+6\beta)^2\omega^2/81}, \quad C = -A\frac{\omega^2-\omega_0^2}{\left(\omega^2-\omega_0^2\right)^2 + 4(\alpha+6\beta)^2\omega^2/81}. \tag{19}$$

Present now the forced solution in the equivalent form

$$\zeta(\theta) = B\cos\omega\theta + C\sin\omega\theta = Amp(\omega_0,\omega)\sin(\omega\theta+\varphi), \tag{20}$$

where the amplitude of forced oscillations is

$$Amp(\omega_0,\omega) = \sqrt{B^2+C^2} = \frac{A}{\sqrt{\left(\omega^2-\omega_0^2\right)^2 + 4(\alpha+6\beta)^2\omega^2/81}}, \tag{21}$$

and the phase $\varphi$ is determined from the equation

$$\tan\varphi = \frac{B}{C} = \frac{2\omega}{9}\frac{\alpha+6\beta}{\omega^2-\omega_0^2}. \tag{22}$$

These expressions can be presented in the traditional dimensionless forms, if we normalise the frequency $\nu = \omega/\omega_0$ and introduce the quality factors, $Q_\alpha = 3\omega_0/2\alpha$ and $Q_\beta = 3\omega_0/4\beta$. Normalising then the amplitude of particle oscillation $A_n = \omega_0^2 Amp/A$, we finally present the amplitude-frequency dependence (21) in the form

$$A_n(\nu, Q_\alpha, Q_\beta) = \frac{1}{\sqrt{(1-\nu^2)^2 + \nu^2 (1/Q_\alpha + 1/Q_\beta)^2}}. \tag{23}$$

And the phase-frequency dependence (22) in the dimensionless form reads

$$\varphi(\nu, Q_\alpha, Q_\beta) = \left(\frac{1}{Q_\alpha} + \frac{1}{Q_\beta}\right) \frac{\nu}{\nu^2 - 1}. \tag{24}$$

The maximum of the resonance curve occurs at $\nu_m = \sqrt{1 - (Q_\alpha + Q_\beta)^2 / 2Q_\alpha^2 Q_\beta^2}$ and equals to

$$(A_n)_{max} = \frac{2Q_\alpha^2 Q_\beta^2}{(Q_\alpha + Q_\beta)\sqrt{4Q_\alpha^2 Q_\beta^2 - (Q_\alpha + Q_\beta)^2}}. \tag{25}$$

These expressions reduce to the classic ones[11,12], when $Q_\beta \to \infty$: $\nu_m|_{Q_\beta \to \infty} = \sqrt{1 - 1/2Q_\alpha^2}$,

$(A_n)_{max}|_{Q_\beta \to \infty} = \dfrac{Q_\alpha}{\sqrt{1 - 1/4Q_\alpha^2}}$. The ratio of maxima is

$$\frac{(A_n)_{max}|_{Q_\beta \to \infty}}{(A_n)_{max}} = \frac{Q_\alpha + Q_\beta}{Q_\beta^2} \sqrt{\frac{4Q_\alpha^2 Q_\beta^2 - (Q_\alpha + Q_\beta)^2}{4Q_\alpha^2 - 1}}. \tag{26}$$

For $\omega_0 = 100$ we obtain $Q_\alpha = 150$ and $Q_\beta = 75$; therefore the maxima ratio as per Eq. (26) is 3. Graphics of dependences of $A_n(\nu, Q_\alpha, Q_\beta)$ and $\varphi(\nu, Q_\alpha, Q_\beta)$ as functions of the normalised frequency $\nu$ are shown in Figs. 2a) and 2b), respectively. Lines 1 pertain to the case when only the HRD force is taken into account, and the MID force is ignored.

Thus, from this analysis one can see that under the influence of MID force the resonance curve of an oscillating bubble becomes three times smaller in amplitude and correspondingly three times wider; the curve maximum is slightly shifted to the left in comparison to the case of HRD force only.

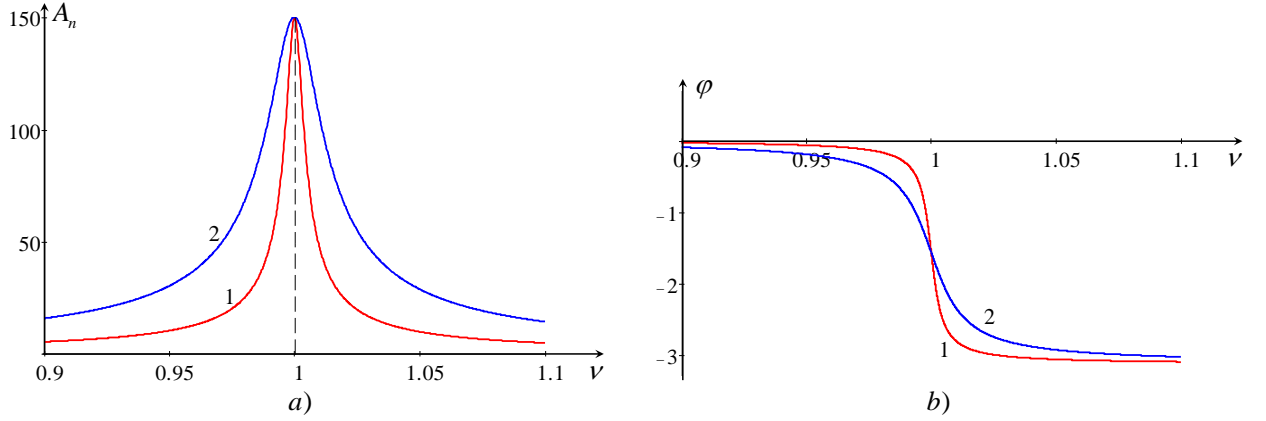

FIG. 2 (color online). The amplitude-frequency dependence (a) and phase-frequency dependence (b) for oscillating bubble in a viscous fluid under the influence of only the HRD force (lines 1) and both HRD and MID forces (lines 2). Function $A_n$ for line 2 in frame (a) was multiplied by the factor 3. The plots were generated for $Q_\alpha = 150$ and $Q_\beta = 75$.

It is of interest to compare the resonant curves of a solid particle of negligible density in comparison with the density of ambient fluid and a gaseous bubble. At first glance these two objects become equivalent, but due to different boundary conditions at the solid-liquid and gas-liquid interfaces, the flow around these objects is different. As a result of that, the dissipation coefficients are also different; these reflects, in particular, in the different coefficients of Stokes and Hadamard–Rybczynski drag forces (cf. Eqs. (3) and (13)) (see, e.g., Ref. [14]). In Fig. 3 we present in the same scale a comparison of resonant characteristics of a solid particle with $r = 0$ (frame a) and a gaseous bubble (frame b). Lines 1 pertain to the cases when the MID forces are ignored, and lines 2 − when these forces are taken into account. As one can see, the resonant curves of a bubble are taller and a bit narrower. In the both cases the influence of MID forces lead to a three–four times reduction of the resonance curves.

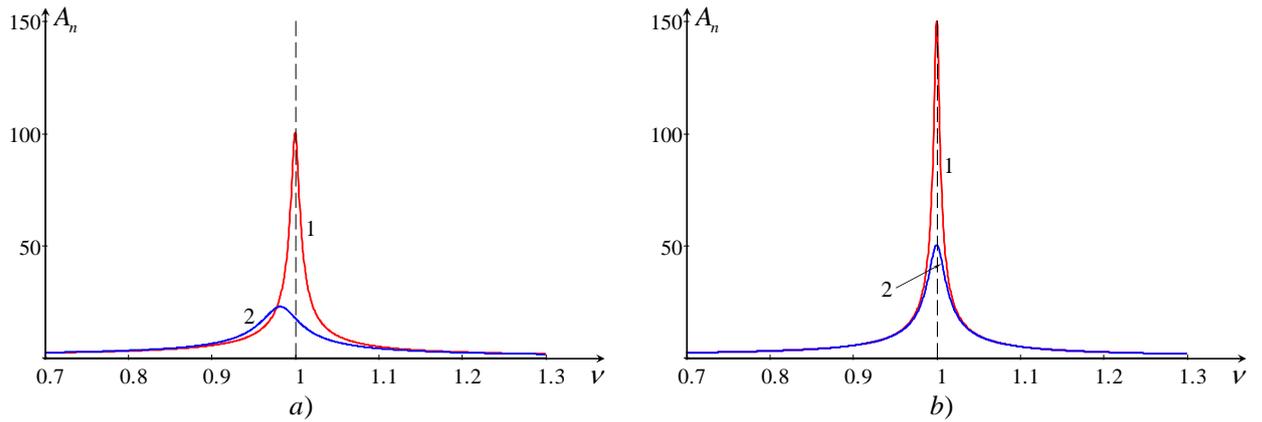

FIG. 3 (color online). The amplitude-frequency dependences of a solid particle of negligible small density (a) gaseous bubble (b) oscillating in a viscous fluid when the MID force is ignored (lines 1) and when it is taken into consideration (lines 2).

Summarising the outcome of this paper, we can conclude that the influence of a MID force on the resonance characteristics of oscillating solid particles and gaseous bubbles in a viscous fluid under the action of external sinusoidal force is significant. The MID force leads to the widening of resonance curves and reduction of resonance peaks which is equivalent, to a certain extent, to the reduction of the quality factor $Q_\alpha$ of the corresponding classical linear oscillator. The results obtained in this study can be helpful in applied problems of micro- and nano-particle and bubble control by the acoustic or electromagnetic fields and, possibly, by other external forces (see, e.g., recent review[10] and references therein).

[1]P. M. Lovalenti, and J. F. Brady, "The force on a Bubble, drop, or particle in arbitrary time-dependent motion at small Reynolds number," Phys. Fluids A. **5**, 2104-2116 (1993).

[2]F. Candelier, J. R. Angilella, and M. Souhar, "On the effect of the Boussinesq-Basset force on the radial migration of Stokes particle in a vortex," Phys. Fluids **16**, 1765-1776 (2004).

[3]F. Candelier, J. R. Angilella, and M. Souhar, "On the effect of inertia and history forces on the slow motion of a spherical solid or gaseous inclusion in a solid-body rotation flow," J. Fluid Mech. **545**, 113-139 (2005).

[4]M. H. Kobayashi, and C. F. M. Coimbra, "On the stability of the Maxey-Riley equation in nonuniform linear flows," Phys. Fluids **17**,113301 (2005).

[5]Y. A. Stepanyants, and G. H. Yeoh, "Particle and bubble dynamics in a creeping flow," Eur. J. Mech. –B/Fluids **28**,619-629 (2009).

[6]Ye. V. Visitskii, A. G. Petrov, and M. M. Shunderyuk, "The motion of a particle in a viscous fluid under gravity, vibration and Basset's force," J. Appl. Math. Mech. **73**, 548-557 (2009).

[7]A. V. Aksenov, A. G. Petrov, and M. M. Shunderyuk, "The motion of solid particle in a fluid in a nonlinear ultrasonic standing wave," Doklady Physics **56**(7), 379-384 (2011).

[8]J. H. Xie, and J. Vanneste, "Dynamics of a spherical particle in an acoustic fields: A multiscale approach." Phys. Fluid **26**, 102001 (2014).

[9]H. K. Hassan, L. A. Ostrovsky, and Y. A. Stepanyants, "Particle dynamics in a viscous fluid under the action of acoustic radiation force," Discontinuity, Nonlinearity, and Complexity, **6**(3), (2017).

[10]L. A. Ostrovsky, and Y. A. Stepanyants, "Dynamics of particles and bubblers under the action of acoustic radiation force," in *Chaotic, Fractional, and Complex Dynamics: New Insights and Perspectives,* edited by M. Edelman, M. Elbert, and M. A. F. Sanjuan (Springer, 2017).

[11]W. McC. Siebert, *Circuits, Signals, and Systems* (MIT Press, 1986).

[12]D. Klepper, and R. Kolenkow, *An Introduction to Mechanics* (Cambridge University Press, 2014).

[13]Y. A. Stepanyants, and G. H. Yeoh, "Nonoparticle dynamics in a viscous fluid at small Reynolds numbers." *Proc. 6$^{th}$ Australasian congress on Applied Mechanics,* (ACAM 6, 12-15 December 2010, Perth, Australia.

[14]L. D. Landau, and E. M. Lifshitz, *Hydrodynamics* (4$^{th}$ ed. Nauka, Moscow(1988)), Engl. Transl.: *Fluid Mechanics* (Pergamon Press, Oxford (1993)).

[15]E. W. Weisstein, *In CRC Concise Encyclopedia of Mathematics*, 2 ed. Chapman & Hall/CRC, Boca Raton, 2003; see also: http://functions.wolfram.com/.